\documentstyle[aaspp4,flushrt]{article}
\input epsf.sty
\let\simlt=\lesssim
\let\simgt=\gtrsim
\begin{document}

\title{Baryonic Features in the Matter Transfer Function}

\author{Daniel J. Eisenstein and Wayne Hu\footnote{Alfred P. Sloan Fellow}}
\affil{Institute for Advanced Study, Princeton, NJ 08540}

\begin{abstract}
\rightskip=0in
We provide scaling relations and fitting formulae for adiabatic
cold dark matter cosmologies that account for all
baryon effects in the matter transfer function to better than $10\%$ in
the large-scale structure regime.  They are based upon a
physically well-motivated separation of the effects of acoustic
oscillations, Compton drag, velocity overshoot,  baryon infall,
adiabatic damping, Silk damping, and cold-dark-matter growth
suppression.  We also find a simpler, more accurate, and better
motivated form for the zero baryon transfer function than previous
works.  These descriptions are employed to quantify the amplitude and
location of baryonic features in linear theory.  While baryonic
oscillations are prominent if the baryon fraction $\Omega_b/\Omega_0
\simgt \Omega_0 h^2 + 0.2$, the main effect in more conventional
cosmologies is a sharp suppression in the transfer function below the
sound horizon.
We provide a simple but accurate description of this effect and
stress that it is {\it not} well approximated by a change in
the shape parameter $\Gamma$.
\end{abstract}
\noindent
\bigskip

\section{Introduction}

A key success of the cold dark matter (CDM) paradigm is the ability of
linear perturbation theory in the early universe to explain the power
spectra observed in the cosmic microwave background
(CMB) and galaxy surveys.  
On the largest scales, {\it COBE} finds a trend of power with
scale rather close to the theoretically motivated scale-invariant
spectrum (\cite{Ben96}\ 1996).  On scales between 10 and 200 Mpc,
however, galaxy surveys (e.g.\ \cite{Bau93}\ 1993; \cite{Str95}\ 1995)
find a much different trend, in which the power increases
with scale.  Merely by including the effects of the transition between
a radiation-dominated and matter-dominated universe, the CDM cosmology
roughly explains both the relative normalization and the differing
spectral indices of these two regimes.

Although the presence of cold dark matter does play a leading role in
determining the matter power spectrum, the inclusion of baryons 
can lead to significant alterations.   Indeed baryonic features in the power
spectrum are a fundamental prediction of the gravitational instability
paradigm, and their discovery would represent a strong consistency test
for the cosmological model.  Such features are the direct result of
small density fluctuations in the early universe prior to recombination.
At those times, the baryons are tightly coupled with the photons and share in 
the same pressure-induced oscillations that lead to acoustic peaks in
the CMB.  This not only leads to intermediate-scale oscillations in the
power spectrum but also produces an overall suppression of power on
small and intermediate scales.

While the low baryon fraction ($\sim\!5\%$) in the standard CDM model
may have justified the neglect of baryonic effects on the power
spectrum in the past,  recent observations favor higher baryon
fractions.  X--ray observations of clusters of galaxies yield baryon
fractions of 10--30\% (\cite{Whi93}\ 1993; \cite{Dav95}\ 1995).
Recent measurements of high-redshift deuterium abundances (\cite{Tyt96}
1996; but see \cite{Rug96}\ 1996) and new theoretical interpretations of
the Lyman--$\alpha$ forest (\cite{Wei97}\ 1997, and references therein)
suggest a value of the baryon density $\Omega_b$ greater than
the fiducial nucleosynthesis value of $0.0125h^{-2}$ (\cite{Wal91}\ 1991).
Meanwhile, observations of large-scale structure
(c.f.\ \cite{Bah96}\ 1996), higher Hubble constants (c.f.\ \cite{Fre96}
1996), and high-redshift galaxies (e.g.\ \cite{Spi97}\ 1996) and
clusters (\cite{Bah97} 1997) favor a universe with $\Omega_0<1$.  Such
baryon fractions lead to modifications of the pure CDM transfer
function that are within observational sensitivities (\cite{Teg97}
1997; \cite{Gol97}\ 1997).

Numerical codes to solve the multi-species Boltzmann equations
(e.g.~\cite{Bon84}\ 1984; \cite{Hol89}\ 1989; \cite{Hu95}\ 1995;
\cite{Sel96}\ 1996) now agree to 1\% accuracy and run in a few minutes
on today's workstations.  While these codes (e.g.\ the publically
available CMBfast) should be used for applications demanding high
accuracy, analytic descriptions are useful for understanding how the
different physical effects give rise to the behavior seen in the
transfer function.  Such descriptions better isolate the unique and
robust observational signatures of physical processes in the early
universe and quantify their scalings with cosmological parameters.
They also probe possible parameter degeneracies and suggest possible
consistency tests with related effects in the CMB.

To this end, we develop here a fitting formula for the matter transfer
function of the general CDM plus baryon universe 
[c.f.\ eqs.\ (\ref{eq:mixed})--(\ref{eq:betab}) and \S 2].  
The formula is composed of a number of well-motivated ingredients,
whose behaviors we discuss in detail.  We achieve fractional accuracies
of 10\% in fully-baryonic models and $\sim\!5\%$ in partial baryon
models.  Included here is a quantification of the fundamental scales
including the acoustic and Silk damping scales which are related but
{\it not} equal to the equivalent scales in the CMB.

We then use this form to produce quantitative assessments of the
amplitude and location of baryonic oscillations as well as the
alteration to the intermediate-scale shape and small-scale
normalization of the transfer function.  Previous assessments of the
latter effects (see e.g.\ \cite{Pea94}\ 1994; \cite{Sug95}\ 1995), though
sufficient for the current generation of experiments at low baryon
fraction, are not accurate enough for future high-precision tests
expected of the Sloan Digital Sky Survey (\cite{Gun95}\ 1995) and 2-Degree
Field survey (\cite{Tay95}\ 1995).  
We present an approximate form that neglects the acoustic
oscillations but accurately represents the suppression of
power on intermediate scales [c.f.\ eqs.\ (\ref{eq:sfit}) and 
(\ref{eq:Gamma})--(\ref{eq:alphagamma})].
To this end, we also present a simpler and more
accurate formula [eq.\ (\ref{eq:T0})] for the zero-baryon transfer function
(e.g.~\cite{Bar86}\ 1986).

In \S 2, we lay the groundwork for the subsequent discussions by
presenting a summary of the physical scales that enter linear
perturbation theory and the exact small-scale solution (Hu \&
Sugiyama\ 1996, hereafter \cite{Hu96}) that we use to anchor our
fitting formula.  In \S 3, we state the fitting formula and discuss its
performance.  In \S 4, we describe the phenomenology revealed by the
formula and present simple scalings to characterize the baryon
oscillations and shape alteration.  We conclude in \S 5.  In Appendix A,
we give a short guide to help the reader locate relevant formulae
from the paper and turn them into COBE-normalized power spectra.
A list of symbols used in this paper is given in Table \ref{tab:symbols}.

\begin{table*}[t]
\begin{center}
\begin{tabular}{|c|l|c|} \tableline
Symbol & Description & Equation \\ \tableline
$\Gamma_0$ & $\Omega_0 h$ & (\ref{eq:Gamma}) \\
$\Gamma_{\rm eff}$ & Effective shape & (\ref{eq:gammaeff}) \\
$\Theta_{2.7}$ & Temperature of CMB & \S \ref{sec:physical} \\
$\alpha_\Gamma$ & $\Gamma$ suppression & (\ref{eq:alphagamma}) \\
$\alpha_b$ & Baryon suppression & (\ref{eq:alphab}) \\
$\alpha_c$ & CDM suppression & (\ref{eq:alphac}) \\
$\beta_b$ & Baryon envelope shift & (\ref{eq:betab}) \\
$\beta_c$ & Shift in CDM log & (\ref{eq:betac}) \\
$\beta_{\rm node}$ & Node shift parameter & (\ref{eq:betak}) \\ 
$R$ & Photon-baryon ratio & (\ref{eq:R}) \\
$R_d$ & $R$ at $z_d$ & (\ref{eq:rsound}) \\
$T$ & Transfer function (TF) & (\ref{eq:Tdef}), (\ref{eq:mixed}) \\
$T_b$ & Baryon sector TF & (\ref{eq:Tbaryonsmall}), (\ref{eq:Tbaryon}) \\
$T_c$ & CDM sector TF & (\ref{eq:Tcdmsmall}), (\ref{eq:Tc}) \\
$T_0$ & Zero baryon TF & (\ref{eq:T0}) \\
$\tilde{T}_0$ & Pressureless TF & (\ref{eq:Tcdmonly}) \\
$k$ & Wavenumber & (\ref{eq:Tdef}) \\
$k_{\rm eq}$ & Equality wavenumber & (\ref{eq:keq}) \\
$k_{\rm Silk}$ & Silk wavenumber & (\ref{eq:ksilk}) \\
$q$ & $k$ scaled with $k_{\rm eq}$ & (\ref{eq:q}) \\
$s$ & Sound horizon at drag epoch & (\ref{eq:rsound}), (\ref{eq:sfit}) \\
$\tilde s$ & Effective sound horizon & (\ref{eq:stilde}) \\
$z_d$ & Redshift of drag epoch & (\ref{eq:zdrag}) \\
$z_{\rm eq}$ & Redshift of equality & (\ref{eq:zeq}) \\
\tableline
\end{tabular}
\end{center}
\caption{A list of symbols used in this paper.}
\label{tab:symbols}
\end{table*}

\section{Physical Effects}
\label{sec:physical}

To motivate and explain the transfer function formulae
in \S \ref{sec:fitting}, we begin with a review of some of the basic 
results of linear perturbation theory, starting with a summary of the
physical scales that enter the theory.  We then describe the exact
small-scale solutions found by \cite{Hu96}, as they play a central role
in the development of the fitting formulae and the explanation of 
baryon phenomenology.

The particular physical properties of the constituents of the universe,
in particular their equations of state and interactions,
can alter the predictions of perturbation theory.
Causality, however, precludes effects at arbitrarily
large scales.  It is therefore usual to measure the resulting perturbations
by comparing them to the amplitude they would have had were causal physics
neglected.  The result is the transfer function, defined as\footnote{in
the synchronous or comoving gauge; gauge issues of are no practical
concern for subhorizon scales.}
\begin{equation}\label{eq:Tdef}
T(k) \equiv {\delta(k,z=0)
	\over \delta(k,z=\infty)} {
	\delta(0,z=\infty) \over \delta(0,z=0)},
\end{equation}
where $\delta(k,z)$ is the density perturbation for wavenumber $k$ and 
redshift $z$.  By construction, $T\rightarrow1$ as $k\rightarrow0$.
The power spectrum $\left<|\delta({\bf k})|^2\right>$ is proportional to
the square of the transfer function multiplied by the initial
power spectrum, most often taken to be proportional to a power law
$k^n$ with $n\approx1$.  Strictly speaking, each species of particle
has a separate transfer function; however, after recombination the
baryons are essentially pressureless and quickly catch up with the cold
dark matter perturbations, leaving both with the same transfer
function.  It is this transfer function that we study in this paper.

We consider cosmologies in which the universe is primarily composed of
photons, baryons (and their accompanying electrons), massless
neutrinos, and cold dark matter (CDM).  Relative to the critical
density, the densities today of the baryons and CDM are $\Omega_b$ and
$\Omega_c$ respectively.  The total matter density is then
$\Omega_0=\Omega_b+\Omega_c$.  The CMB temperature $T_{\rm CMB}$
is written as $2.7\Theta_{2.7}{\rm\,K}$; the best observations from the
{\it COBE} FIRAS instrument are $2.728 \pm 0.004 {\rm \, K}$
(\cite{Fix96}\ 1996; 95\% confidence interval).  We assume that the
massless neutrinos contribute an energy density corresponding to three
species at $(4/11)^{1/3}$ the temperature of the photons.  We use a
Hubble constant $H_0$ and define $h\equiv H_0/(100 {\rm
km\,s^{-1}\,Mpc^{-1}})$.  It is important to note that for $z \gg
\Omega_0^{-1}$ the dynamics of the fluctuations are determined solely by the
matter-radiation ratio $\Omega_0 h^2 \Theta_{2.7}^{-4}$, the
baryon-photon ratio $\Omega_b h^2 \Theta_{2.7}^{-4}$, and the
CMB temperature $T_{\rm CMB}$.  We fix the last of these
to be the {\it COBE} value and do not include variations in
$\Theta_{2.7}$ in our fits.  Since all effects in the transfer function
are set at those early times, the resulting description should depend
only on $\Omega_0 h^2$ and $\Omega_b/\Omega_0$.  
The existence today of a non-zero
cosmological constant or curvature is insignificant.

\subsection{Length and Time Scales}
\label{sec:scales}

The physics governing the evolution of perturbations in CDM--baryon
universes involves three distinct length scales: the horizon size at
matter-radiation equality, the sound horizon at the time of recombination,
and the Silk damping length at recombination.

In the usual cosmological paradigm, non-relativistic particles (baryons, 
electrons, and CDM) dominate relativistic particles (photons and
massless neutrinos) in density today.  However, because the density of these two classes of
particles scale differently in time, at an earlier time, the 
reverse situation held.  The transition from a radiation-dominated
universe to a matter-dominated one occurs roughly at 
\begin{equation}\label{eq:zeq}
z_{\rm eq} = 2.50\times10^4 \Omega_0 h^2 \Theta_{2.7}^{-4},
\end{equation}
the redshift where the two classes have equal density. 
As density perturbations behave differently in a radiation-dominated
universe versus a matter-dominated one due to pressure support, the scale of 
the particle horizon at the equality epoch $z_{\rm eq}$,
\begin{equation}\label{eq:keq}
k_{\rm eq} \equiv (2 \Omega_0 H_0^2 z_{\rm eq})^{1/2} 
       = 7.46\times10^{-2} \Omega_0 h^2 \Theta_{2.7}^{-2} {\rm\,Mpc^{-1}},
\end{equation}
is imprinted on the matter transfer function;
in particular, perturbations on smaller scales are suppressed in 
amplitude in comparison to those on large scales.
If the universe consisted only of non-interacting matter and radiation,
the matter transfer function would depend on the ratio $(k/k_{\rm eq})$ alone. 

Complications arise due to interactions between the species.  
Prior to the recombination of baryons and electrons,
the large density of free electrons couples the baryons 
to the photons through Coulomb and Compton interactions so that the 
three species move together as a single fluid.  
This continues until, in the process of recombination, the rate
of Compton scattering between photons and electrons becomes too low,
freeing the baryons from the photons. We thus define the
drag epoch $z_d$ as the time at which the baryons are released from the
Compton drag of the photons in terms of a weighted integral over the
Thomson scattering rate (see \cite{Hu96}, Eqn. C8, E2).  A fit to the numerical
recombination results is 
\begin{eqnarray}
z_d & = & 1291 {(\Omega_0 h^2)^{0.251}
\over 1 + 0.659 (\Omega_0 h^2)^{0.828} }
[1 + b_1 (\Omega_b h^2)^{b_2}], \nonumber\\
b_1 & = & 0.313 (\Omega_0 h^2)^{-0.419} [1 + 0.607 
(\Omega_0 h^2)^{0.674} ], \nonumber\\
b_2 & = & 0.238 (\Omega_0 h^2)^{0.223},  \label{eq:zdrag}
\end{eqnarray}
where we have reduced $z_d$ by a factor of $0.96$ from \cite{Hu96} 
on phenomenological grounds.   For $\Omega_b h^2 \simlt 0.03$,
this epoch follows last scattering of the photons.

\begin{figure}[bt]
\centerline{\epsfxsize=\textwidth \epsffile{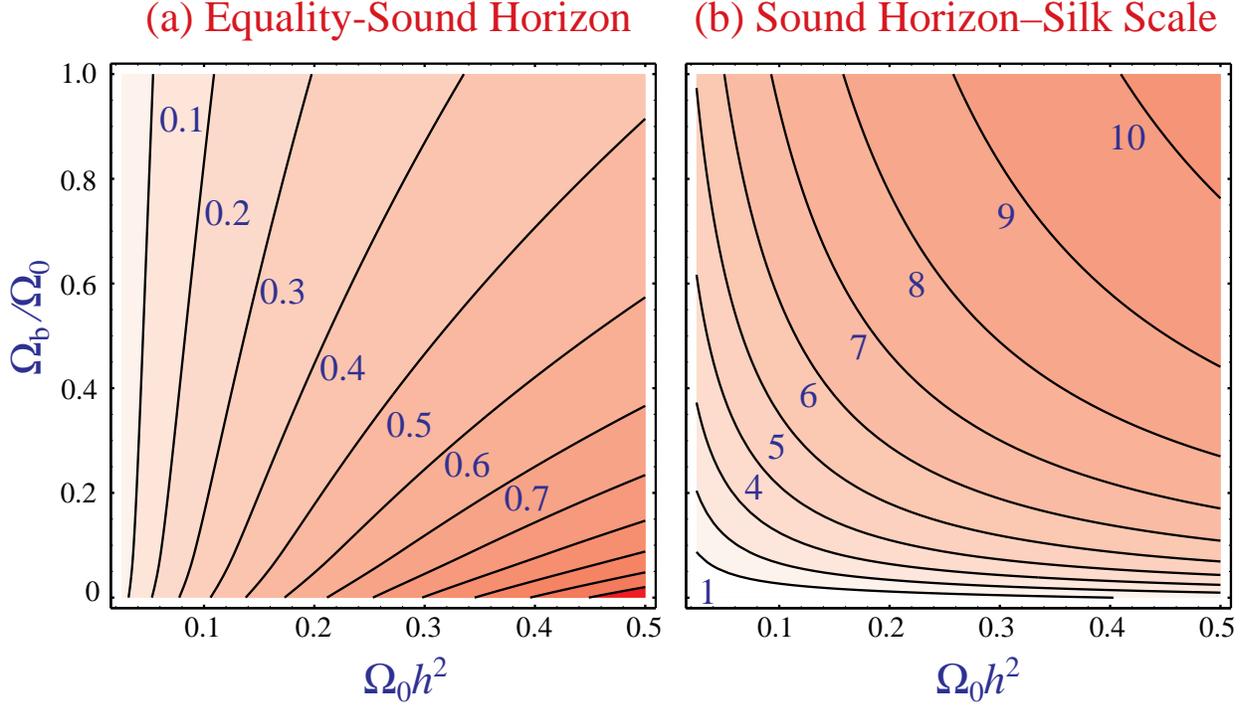}}
\caption{Comparison of the physical scales as functions of $\Omega_0 h^2$ 
and the baryon fraction $\Omega_b/\Omega_0$.  
(a) The equality scale vs.\ the sound horizon: $k_{\rm eq}s/\pi$ 
(unlabeled contours at $0.1$ increments).  
(b) The sound horizon vs.\ the Silk scale: $k_{\rm Silk}s/\pi$ 
(unlabeled contours $2$ and $3$). 
The factors of $\pi$ have been included to facilitate comparison with 
the acoustic scale.}
\label{fig:scales}
\end{figure}

Prior to $z_d$, small-scale perturbations in the photon-baryon fluid
propagate as acoustic waves.  The sound speed is $c_s = 1/\sqrt{3(1+R)}$
(in units where the speed of light is unity), where $R$ is the
ratio of the baryon to photon momentum density,
\begin{equation} \label{eq:R}
R  \equiv 3\rho_b/4\rho_\gamma = 
31.5 \Omega_b h^2 \Theta_{2.7}^{-4} (z/10^3)^{-1}.
\end{equation}
We define the sound horizon at the drag epoch as the comoving distance a wave
can travel prior to redshift $z_d$,
\begin{equation} \label{eq:rsound}
s = \int_0^{t(z_d)}\,c_s\, (1+z)dt
	= {2\over3k_{\rm eq}} \sqrt{6\over R_{\rm eq}}
	\ln { \sqrt{ 1 + R_d} + \sqrt{ R_d + R_{\rm eq}}
        \over 1 + \sqrt{R_{\rm eq}} },
\end{equation}
where $R_d\equiv R(z_d)$ and $R_{\rm eq}\equiv R(z_{\rm eq})$ are the values 
of $R$ at the drag epoch and epoch of matter-radiation equality, respectively.  
The sound horizon at the drag epoch (hereafter simply the sound horizon)
is larger than the equality horizon ($\sim\!1/k_{\rm eq}$) in
high-$\Omega_0$ models but smaller than it in low-$\Omega_0$ models;
it also decreases strongly with increasing baryon fraction if 
$R_d \simgt 1$ (see Fig.~\ref{fig:scales}).

On small scales, the coupling between the baryons and the photons is 
not perfect, such that the two species are able to diffuse past 
one another (\cite{Sil68}\ 1968).  The Silk damping scale is well fit 
by the approximation
\begin{equation} \label{eq:ksilk}
k_{\rm Silk} = 1.6 (\Omega_b h^2)^{0.52} (\Omega_0 h^2)^{0.73} 
\left[1+(10.4\Omega_0 h^2)^{-0.95}\right] {\rm\,Mpc}^{-1},
\end{equation} 
which represents a $\pm 20\%$ phenomenological correction from 
the value given in \cite{Hu96}.  The Silk scale is generally a smaller
length scale than either $s$ or $1/k_{\rm eq}$.  Note that the difference
between the drag and last scattering epochs implies that for 
$\Omega_b h^2 \simlt 0.03$
the sound and Silk scales in the transfer function are larger than 
those in the CMB.

We show a comparison of these scales as a function of cosmological
parameters in Figure \ref{fig:scales}.

\subsection{Small-scale Solutions}
\label{sec:small}

In the small-scale limit, one can solve the linear perturbation equations
analytically in the approximation that baryons provide no gravitational 
source to the CDM (\cite{Hu96}).  This approximation is appropriate below the
sound horizon since baryon perturbations are pressure supported. 
As we will use this solution in order to anchor the small-scale end of
our fitting formulae, we describe the solutions further.

\begin{figure}[bt]
\centerline{\epsfxsize=\textwidth \epsffile{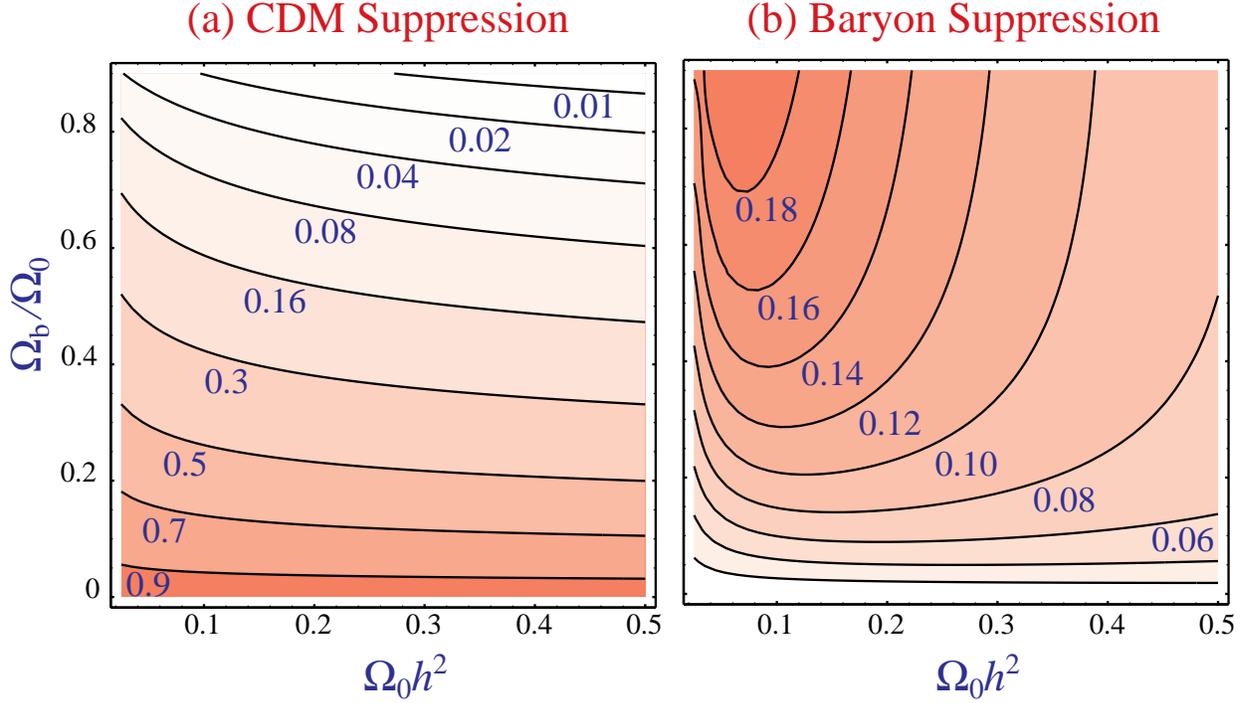}}
\caption{Suppression factors for (a) the CDM ($\alpha_c \Omega_c/\Omega_0$)
and (b) the baryonic acoustic oscillations ($\alpha_b \Omega_b/\Omega_0$).}
\label{fig:suppression}
\end{figure}

The transfer function is written as a sum
of the baryon and cold dark matter contributions at the drag epoch
\begin{equation}
T(k) = {\Omega_b \over \Omega_0} T_b(k) + 
       {\Omega_c \over \Omega_0} T_c(k).
\end{equation}
The CDM transfer function can be solved exactly
in terms of hypergeometric functions that are more conveniently approximated
by the following form:
\begin{eqnarray}\label{eq:Tcdmsmall}
T_{c} &\rightarrow& \alpha_c {\ln 1.8\beta_c q\over 14.2q^2}, \\
q &=& (k/{\rm\,Mpc}^{-1})\Theta_{2.7}^2 (\Omega_0 h^2)^{-1} 
= {k\over 13.41 k_{\rm eq}},\label{eq:q}
\end{eqnarray}
where $\alpha_c$ and $\beta_c$ are fit by
\begin{eqnarray} \label{eq:alphac}
\alpha_c & =& a_1^{-\Omega_b/\Omega_0} a_2^{-(\Omega_b/\Omega_0)^3}, \\
a_1 &=& (46.9\Omega_0 h^2)^{0.670} [1+(32.1 \Omega_0 h^2)^{-0.532}], \nonumber\\
a_2 &=& (12.0\Omega_0 h^2 )^{0.424} [1+(45.0 \Omega_0 h^2)^{-0.582}], 
\nonumber\\[3pt]
\beta_c^{-1} & =& 1 + b_1 [(\Omega_c/\Omega_0)^{b_2}-1], \label{eq:betac}\\
b_1 & = &0.944 [ 1 + (458 \Omega_0 h^2)^{-0.708} ]^{-1}, \nonumber\\
b_2 & = &(0.395 \Omega_0 h^2)^{-0.0266}.  \nonumber
\end{eqnarray}
As $\Omega_b/\Omega_0 \rightarrow 0$, $\alpha_c, \beta_c\rightarrow1$.
Equation (\ref{eq:Tcdmsmall}) shows the familiar $\ln(k)/k^2$ dependence of
the small-scale CDM transfer function. This occurs because 
outside the horizon, density perturbations grow as $k^2$
due the product of potential and velocity gradients which drive the
growth; inside the horizon in the radiation dominated epoch
the growth is logarithmic.
The main effect of the baryons comes
from the suppression in growth rates between equality and the drag
epoch.  As $\Omega_0 h^2$ increases the time between the two epochs
increases; thus the maximum suppression due to the baryons occurs in
the highest $\Omega_0 h^2$ models.  A plot of $\alpha_c
\Omega_c/\Omega_0$ is shown in Figure \ref{fig:suppression}a.

In the small-scale limit, the baryons are trapped in acoustic oscillations
until recombination permits them to slip past the photons.  
While the density perturbation of this oscillation contributes to the
transfer function, the corresponding velocity perturbation actually
dominates in the small-scale limit.  When the oscillation is released
at the drag epoch, the baryons move kinematically according to their 
velocity and generate a new density perturbation (\cite{Sun70}\ 1970;
\cite{Pre80}\ 1980).  This so-called 
velocity overshoot means that the transfer function for $ks\gg1$ follows
\begin{equation}\label{eq:Tbaryonsmall}
T_b \rightarrow \alpha_b {\sin(ks)\over ks} {\cal D}(k).
\end{equation}
Here ${\cal D}(k)$ represents the effects of Silk damping, which occurs 
due to combination of diffusion of
the photons with respect to the baryons and Compton drag moving baryons from
overdensities to underdensities and hence destroying the perturbation.
That the dependence is $\sin(ks)$ rather than $\cos(ks)$ is the result
of the dominance of the velocity term rather than the density term.
A detailed treatment allows one to calculate $\alpha_b$:
\begin{eqnarray}\label{eq:alphab}
\alpha_b &=& 
2.07 k_{\rm eq}s (1+R_d)^{-3/4} G\left({1+z_{\rm eq} \over 1+z_d }\right)\, ,\\
G(y)& = & y\left[-6\sqrt{1+y}+(2+3y)\ln\left(
\sqrt{1+y}+1\over\sqrt{1+y}-1\right)\right],
\end{eqnarray}
The factor $(1+R_d)^{-3/4}$ comes from the damping of
oscillations resulting from the adiabatic decrease in the 
sound speed (\cite{Pee70}\ 1970; \cite{Hu96}, eq.\ [A17]).
The factor involving $G(y)$ $(\propto y^{-1/2}$ for $y\gg 1$)
comes from the product of the growth suppression between equality and the
drag epoch $(\propto y^{-1})$ and the time available before the
velocities creating the perturbation decay due to the free expansion of
the universe $(\propto y^{1/2})$.  A plot of $\alpha_b
\Omega_b/\Omega_0$ is shown in Figure \ref{fig:suppression}b.

That the phase of the oscillations is $ks$ (\cite{Pee70}\ 1970) can be
seen simply from integrating the dispersion relation $\omega= kc_s$.  
The change in
phase for an acoustic wave is $\delta\phi=[k(1+z)] c_s \delta t$.
Integrating this from $t\approx0$ to the drag epoch (where the baryons
are released and the oscillations freeze-out) yields $ks$, owing to the
definition in equation (\ref{eq:rsound}).  A technical complication occurs for
$k\simgt k_{\rm Silk}$.  The presence of strong damping slightly raises
the redshift at which the oscillations freeze-out, making $s$ a few
percent smaller (see \cite{Hu96} Fig.~2).  We have neglected this
effect because it occurs at sufficiently small scales that the
resulting phase shift is unobservable in practice, but one can see the
deviations when comparing to numerical results
(c.f.\ Fig.~\ref{fig:poster}).

\section{Fitting Formulae}
\label{sec:fitting}

As we have seen in \S \ref{sec:physical},
analytic solutions exist for the transfer function at both large and
small scales.  The transition between these extremes is defined by two
scales, the horizon at matter-radiation equality and the sound horizon
at the end of the drag epoch.  The former sets the dynamics of the
expansion and perturbation growth; the latter sets the scale at which
pressure support becomes important for the baryons.  Because the range
of scales accessible by the study of structure formation falls within
this transition regime, it is important to understand the full transfer
function in detail.  To that end, we present in this section a fitting
formula that approximates the full transfer function on all scales.

We write the transfer function as the sum of two pieces,
\begin{equation} \label{eq:mixed} 
T(k) = {\Omega_b \over \Omega_0} T_b(k) + {\Omega_c \over \Omega_0} T_c(k), 
\end{equation} 
whose origins
lie in the evolution before the drag epoch of the baryons and cold dark
matter respectively.  This separation is physically reasonable as
before the drag epoch the two species were dynamically independent and
after the drag epoch their fluctuations are weighted by the fractional
density they contribute.  This automatically includes in $T_c$ the effects of
baryonic infall into CDM potential wells.  Note however that $T_b$ and
$T_c$ are themselves {\it not} true transfer functions as they do not
reflect the density perturbations of the relevant species today.
Rather, it is their density-weighted average $T(k)$ that is the 
transfer function for both the baryons and the CDM.

\subsection{Cold Dark Matter}
\label{sec:Tc}

The transfer function for cosmologies in which non-interacting cold
dark matter dominates over baryons have been studied by many authors
and accurate fitting formulae already exist in this limit
(e.g.\ \cite{Bon84}\ 1984; \cite{Bar86}\ 1986; but see improvements in \S
\ref{sec:shape}) However the effect of baryons, though long known from
numerical calculations (e.g.\ \cite{Pee70}\ 1970; \cite{Hol89}\ 1989),
have in the past either been included in fitting formulae in an ad hoc
manner (see e.g.\ \cite{Pea94}\ 1994; \cite{Sug95}\ 1995) or only in the
small scale limit (\cite{Hu96}).

In the presence of baryons, the growth of CDM perturbations is suppressed on
scales below the sound horizon.  The change to the asymptotic form can
be calculated and has been shown in equations 
(\ref{eq:Tcdmsmall})--(\ref{eq:betac}).  We introduce this
suppression by interpolating between two solutions near the sound horizon:
\begin{eqnarray}\label{eq:Tc}
T_c(k) &=& f \tilde{T}_0(k, 1, \beta_c) + (1-f) \tilde{T}_0(k, \alpha_c, \beta_c)\\
f &=& {1\over 1+(ks/5.4)^4}, \label{eq:fcdm}
\end{eqnarray}
with
\begin{eqnarray} \label{eq:Tcdmonly}
\tilde{T}_0(k,\alpha_c, \beta_c) &=& 
{\ln(e + 1.8\beta_c q) \over \ln(e + 1.8 \beta_c q) + C q^2},\\
C&=& {14.2 \over\alpha_c} + {386 \over 1 + 69.9q^{1.08}}.
\end{eqnarray}
$q$, $\alpha_c$, and $\beta_c$ have been given in
equations (\ref{eq:q}), (\ref{eq:alphac}), and (\ref{eq:betac}), respectively.

\subsection{Baryons}
\label{sec:Tb}

In the case of cosmologies without cold dark matter, the transfer
function departs from unity below the sound horizon to exhibit a series
of declining peaks due to acoustic oscillations.  The small-scale exact
solution of equation (\ref{eq:Tbaryonsmall}) suggests that these may be written
as the product of a declining oscillatory term, a suppression due to
the decay of potentials between the equality and drag scales, and an
exponential Silk damping.  We therefore write
\begin{eqnarray} \label{eq:Tbaryon}
T_b &=& 
\left[{\tilde{T}_0(k; 1,1)\over 1+\left(ks/ 5.2\right)^2}+ 
{\alpha_b \over 1+\left(\beta_b/ks\right)^3} e^{-(k/k_{\rm Silk})^{1.4}}\right] 
j_0(k\tilde s). 
\end{eqnarray}
Here the spherical Bessel function $j_0(x)\equiv \sin(x)/x$ is a 
piece that approaches unity above the sound horizon but oscillates below it.
The envelope in square brackets traces the zero-baryon CDM case above
the sound horizon and then breaks to a constant multiplied by an
exponential Silk damping factor.
We attach the Silk damping factor only to the second term because such
diffusion can only occur on scales below the sound horizon, where only
the second term contributes significantly.  This subtlety marginally
improves the fit.
The sound horizon $s$, Silk scale $k_{\rm Silk}$, and amplitude
suppression $\alpha_b$ were given in equations (\ref{eq:rsound}),
(\ref{eq:ksilk}), and (\ref{eq:alphab}) respectively; we now discuss
$\tilde{s}$ and $\beta_b$.

While the nodes of the baryonic transfer function asymptotically
approach those of $\sin(ks)$ for $ks\gg1$, the first few nodes
($ks\lesssim10$) fall at higher $k$ than predicted by $\sin(ks)$.  This
shift is due to the contribution of the baryon density perturbation
itself at the drag epoch and reflects the fact that at the sound
horizon velocity overshoot is not the dominant effect.  This effect
increases with $\Omega_0 h^2$ because the time available for velocity
overshoot (see eq.\ [\ref{eq:alphab}]) decreases as $(z_d/z_{\rm eq})^{1/2}$
and is only weakly dependent on the baryon fraction.  We have verified
this explanation of the node shift by isolating the density and
velocity contributions of the baryons at the drag epoch from numerical
evolution codes.

We address this shifting of the nodes phenomenologically
by introducing the quantity
\begin{equation}\label{eq:stilde}
\tilde{s}(k) = {s\over \left(1+(\beta_{\rm node}/ks)^3\right)^{1/3}}.
\end{equation}
For $ks\gg\beta_{\rm node}$, $\tilde{s}\rightarrow s$, restoring the
sinusoidal nodes.  However, at $ks\lesssim\beta_{\rm node}$,
$\tilde{s}\approx ks^2/\beta_{\rm node}<s$, moving the nodes to higher
$k$.  We find
\begin{equation}\label{eq:betak}
\beta_{\rm node} = 8.41 (\Omega_0 h^2)^{0.435},
\end{equation}
independent of the baryon fraction.
Hence the effect gets smaller at low $\Omega_0$ as expected.

The amplitude $\alpha_b$ specifies the small-scale asymptotic
contribution of the velocity portion of the acoustic oscillation.  Two
effects modify this amplitude at large scales.  Above the sound
horizon, velocities contributions to the transfer function fall off.  
Furthermore, the amplitude declines if CDM
dominates the energy density of the photon-baryon system when the
wavelength enters the horizon.  This occurs due the absence of feedback
in the gravitational driving of the photon-baryon oscillator
(\cite{Hu96} \S 3.1).  The net result is a cut-off associated with the sound
horizon which moves to smaller scales as $\Omega_0 h^2$ increases
and/or $\Omega_b/\Omega_0$ decreases.  We describe this in
equation (\ref{eq:Tbaryon}) by turning on the velocity term at the
characteristic scale $\beta_b s$, where
\begin{equation}\label{eq:betab}
\beta_b = 0.5+{\Omega_b\over\Omega_0}+(3-2{\Omega_b\over\Omega_0})
\sqrt{(17.2\Omega_0 h^2)^2+1}.
\end{equation}

\subsection{Performance}

\begin{figure}[p]
\centerline{\epsfxsize=\textwidth \epsffile{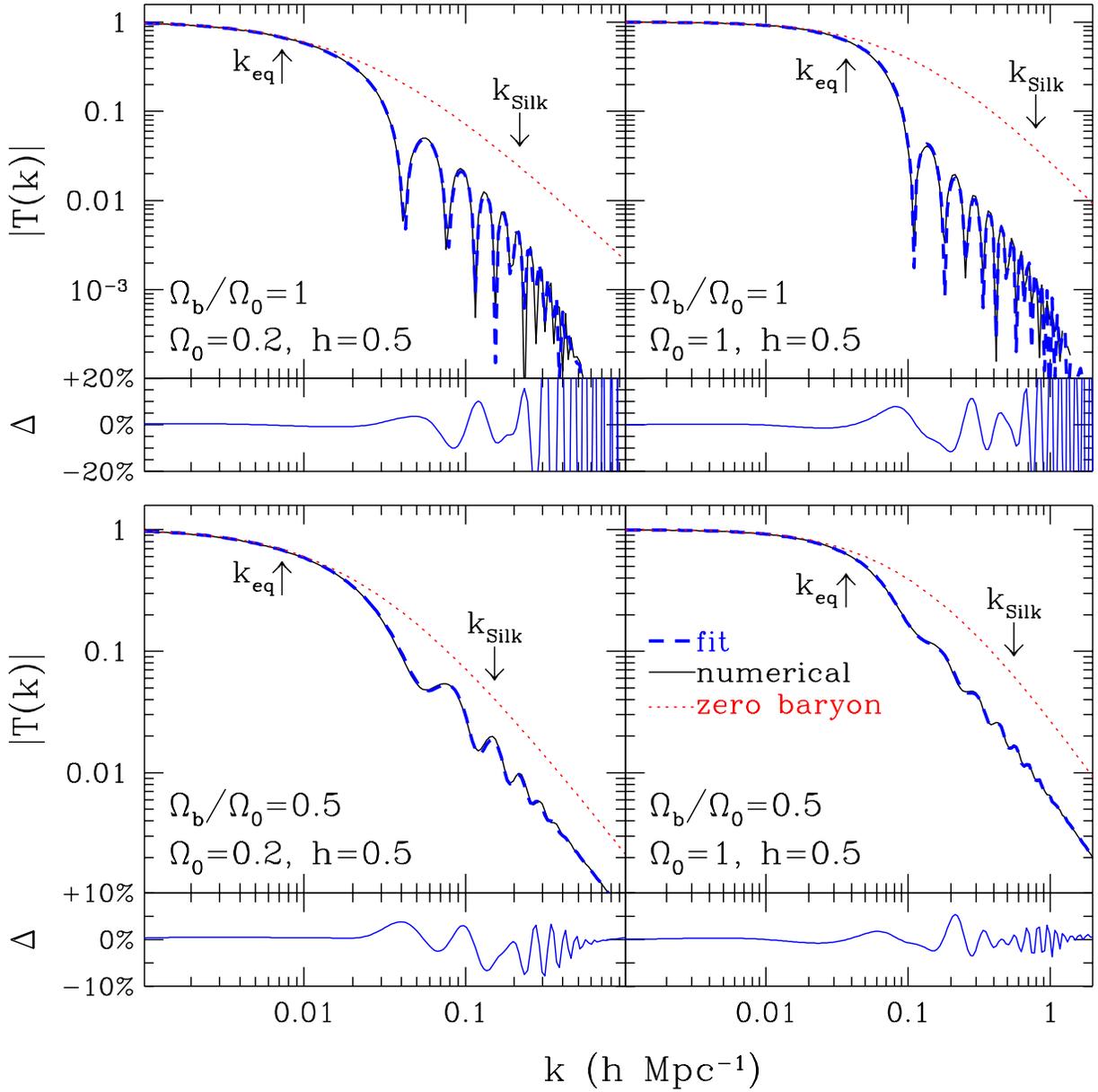}}
\caption{Four examples of the fit compared to numerical results.  The 
larger plots show the numerical result (solid) and the fit (dashed).
The smaller subplots show the residuals, defined as the difference between
the two divided by a non-oscillatory envelope.  Note that in the
fully baryonic models, the oscillations have alternating sign in the
transfer function.  Also shown is the zero baryon case (dotted); note the strong
suppression on scales below the sound horizon due to the baryons.}
\label{fig:poster}
\end{figure}

For the parameter range $0.025\lesssim\Omega_0 h^2 \lesssim 0.25$ 
and $0 \le \Omega_b/\Omega_0 \le 1$ the
fitting formula works quite well.  For fully baryonic models (i.e.\ 
$\Omega_c=0$) the fractional residuals are nearly always under 10\%.
As the baryon fraction decreases, the accuracy improves due to the
increasing contribution of the simpler CDM piece.  Residuals
smaller than 5\% are typical for $\Omega_b/\Omega_0<0.5$.  Note that
we quote the residuals as the difference between the fit and the numerical
results divided by a non-oscillatory envelope that is defined by replacing
$\sin(k\tilde{s})/k\tilde{s}$ in equation (\ref{eq:Tbaryon}) by 
$[1+(k\tilde{s})^4]^{-1/4}$.  This envelope matches the knee of
the transfer function and grazes all the subsequent maxima.

In Figure \ref{fig:poster} we display four comparisons of the fitting
formula relative to the numerical results.  Also shown are the residuals
relative to our envelope.  The reason for the degradation in the fit
at the shortest scales is that small errors in the sound horizon 
(c.f.\ the end of \S \ref{sec:small}) produce significant errors in the
phase of the oscillations, producing order unity residuals.  However,
the fitting formula reproduces the correct amplitude and hence the Silk
scale.

The most serious systematic error in the fitting formula occurs for
$\Omega_0 h^2 \simgt 0.25$.  In the baryon sector of these cases, the 
drop between $ks \approx 1$ and the oscillations at $ks\simgt 5$
becomes quite precipitous.  Our formula does not decline this quickly, 
causing the amplitude of the first valley to be significantly overestimated.
Later peaks and valleys are underestimated in an attempt to compensate.
One can see the beginnings of this trend in the $\Omega_0 = \Omega_b = 1$
example in Figure \ref{fig:poster}; the problem gets more severe for
higher $\Omega_0 h^2$.

A less important systematic effect occurs for high baryon fraction
($0.7<\Omega_b/\Omega_0<0.9$), low-$\Omega$ models.  Due to a small shift
in the CDM break scale (eq.\ [\ref{eq:fcdm}]) that we have chosen not
to model, the first valley is systematically underestimated by
$\sim\!10$--15\% in amplitude.  This problem does not extend to lower
baryon fractions.

\section{Phenomenology}

There are a number of phenomenological trends as a function of 
cosmological parameters.  The two basic effects that arise from the
inclusion of baryons are the introduction of oscillations and the 
suppression of power below the sound horizon, with a corresponding
sharpening of the bend around the sound horizon.  
We discuss these two in turn.

\subsection{Baryon Oscillations}

\begin{figure}[bt]
\centerline{\epsfxsize=3.5in \epsffile{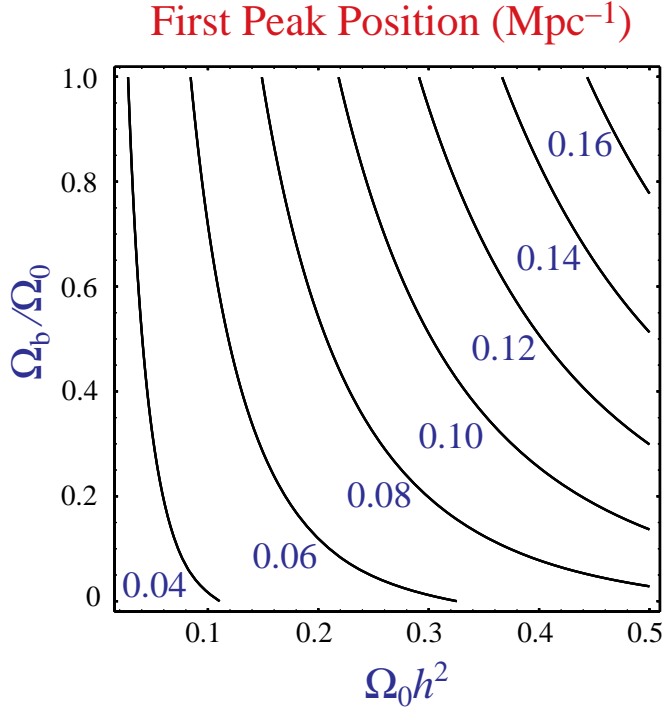}}
\caption{The location of the first peak in Mpc$^{-1}$ as a function
of cosmological parameters.}
\label{fig:bumpplace}
\end{figure}

Two interesting aspects of the baryonic oscillations are the location
and amplitude of the peaks and troughs.  Well under the sound horizon
we expect them to fall at $k = m\pi/2s$ where $m=3,7,11\ldots$ for
troughs and $m=5,9,13\ldots$ for peaks.  However as described in \S
\ref{sec:Tb}, the first few oscillations are shifted according to the
parameter $\beta_{\rm node}$.  As $\Omega_0 h^2$ increases, the first
few peaks and troughs are progressively shifted to higher $k$.  A
corollary to this shift is that the ratio of the node locations becomes
smaller as one raises $\Omega_0$, i.e.\ the valleys and peaks become
slightly narrower.  The location of the first peak is conveniently fit
as
\begin{equation}
k_{\rm peak} = {5 \pi \over 2s} (1+ 0.217 \Omega_0 h^2)
\end{equation}
where
\begin{equation}\label{eq:sfit}
s = {44.5 \, \ln(9.83/\Omega_0h^2) \over \sqrt{1+10(\Omega_bh^2)^{3/4}}}
{\rm\,Mpc}
\end{equation}
approximates the sound horizon to $\sim 2\%$ over the range $\Omega_b
h^2 \simgt 0.0125$ and $0.025\simlt\Omega_0 h^2\simlt 0.5$.
The value of $k_{\rm peak}$ as a function of cosmological parameters is
shown in Fig.~\ref{fig:bumpplace}.

\begin{figure}[bt]
\centerline{\epsfxsize=\textwidth \epsffile{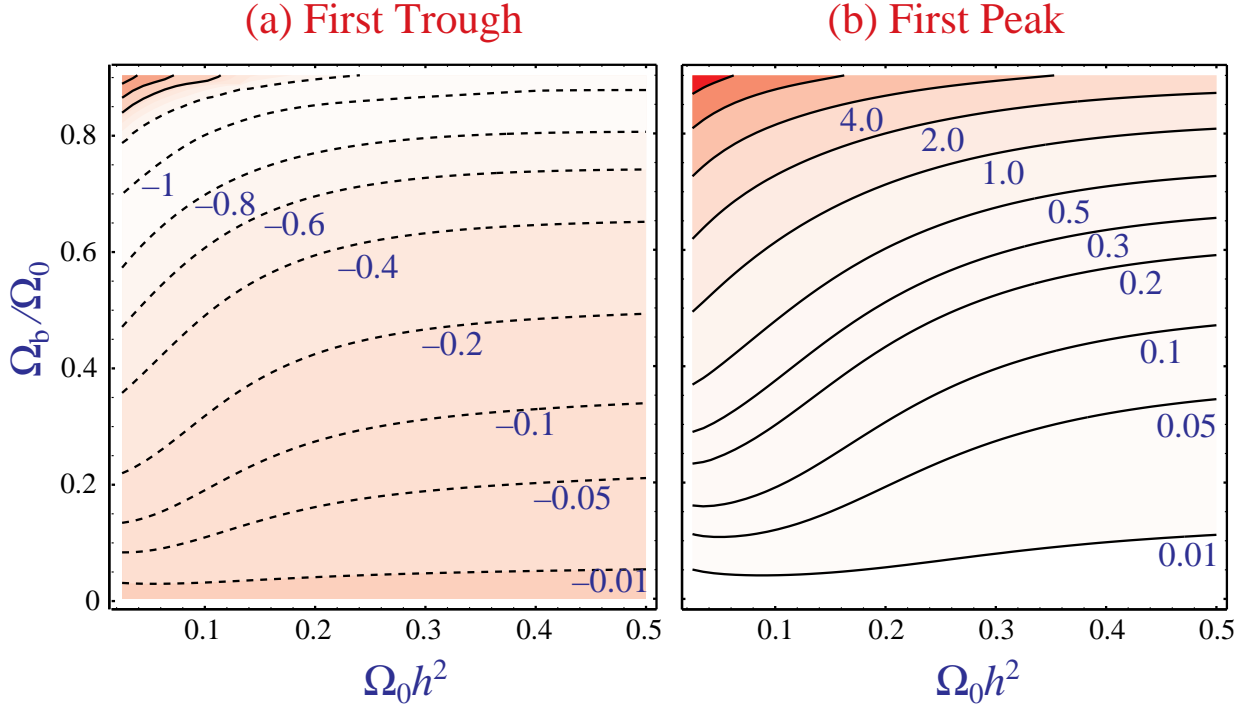}}
\caption{The fractional enhancement of power due to (a) the first valley
and (b) the first peak relative to non-oscillatory CDM portion of the transfer
function, $(T/T_c)^2-1$, at the appropriate wave vector from the fit.
Unlabeled contours are at (a) -0.8, 0, 1, 3  and (b) 8.0, 16.0}
\label{fig:size}
\end{figure}

The amplitude of the oscillations also has a non-trivial dependence on
the cosmological parameters.  The oscillations of course grow stronger
as the baryon fraction increases.  However, at {\it fixed} baryon
fraction, they are weaker compared with CDM contributions in high
$\Omega_0 h^2$ due to the increase in the time available for the CDM to
grow between equality and the drag epoch.  While the full series of
peaks and valleys may be impossible to observe due to non-linear
structure formation, the first valley and peak are generally in the
linear regime.  In Fig.~\ref{fig:size}, we show the fractional
enhancement of power due to the oscillations over the smooth CDM
contributions.  The first peak grows monotonically with the baryon
fraction.  The first trough is more subtle: while the transfer function
at this location simply declines as the baryon fraction increases, when
$T(k)$ goes negative the power, which is the square of $T$, will
regenerate.  Perfect cancellation of the baryon and CDM contributions
occurs along the contour labeled ``-1'' in Fig.~\ref{fig:size}a; above
this line the trough in amplitude becomes a peak in power as the baryon
contributions come to fully dominate.  A useful rough scaling as to
when oscillations become important is given by
\begin{equation}
{\Omega_b \over \Omega_0} \simgt \Omega_0 h^2 + 0.2
\end{equation}
which crudely describes the region where the change in power is greater
than $\sim 20\%$.

A related trend is the increase with $\Omega_0 h^2$ of the sharpness
of the decline in the baryonic sector from the knee at $ks\approx1$ 
($T\approx1$) into the series of oscillations below the sound horizon.
This is most easily seen in the fully baryonic models of Figure 
\ref{fig:poster}; near $k=0.1h {\rm\,Mpc}^{-1}$ in the $\Omega_0 h^2 = 0.25$
model, the transfer function drops a factor of ten in under half a 
decade in $k$.  This break becomes even more striking in higher $\Omega_0
h^2$ cases.

\subsection{Effective Shape}
\label{sec:shape}

As we have seen, if $\Omega_b/\Omega_0 \simlt \Omega_0 h^2 + 0.2$, the
main effect of the baryons is {\it not} to introduce oscillations into
the transfer function but to suppress the $k^{-2}$ tail from the growth
of CDM perturbations.  This occurs both because the CDM portion $T_c$
is suppressed by $\alpha_c$ and because the baryonic portion $T_b$ is
providing essentially no power below the sound horizon.  As noted just
above, the latter transition can occur quite quickly.  These
suppressions indicate that the {\it shape} of the transfer function
must change, with a break near the sound horizon.  It is useful to
quantify this effect.

Let us work forward from the zero baryon case.  Here the transfer
function is parameterized by $q \propto k/k_{\rm eq}$, more commonly
expressed as a shape parameter $\Gamma=\Omega_0 h$, where
\begin{equation}\label{eq:Gamma}
q = {k \over {h \rm\,Mpc}^{-1}}\Theta_{2.7}^2 /\Gamma.
\end{equation}
A commonly used fitting formula to the zero-baryon limit was presented
by \cite{Bar86}\ (1986, eq.\ G3).  However this formula fits neither the exact
small-scale solution of \S \ref{sec:small} nor does it have the
quadratic deviation from unity required by the theory.  The latter is a
fundamental requirement of causality (\cite{Zel65}\ 1965), in that one
power of $k$ must arise from stress gradients generating bulk velocity and
a second from velocity gradients generating density perturbations.  In
fact the coefficient of this quadratic deviation can be calculated 
perturbatively if the stress gradients
are dominated by the isotropic (pressure) term.

\begin{figure}[bt]
\begin{center} \leavevmode \epsfxsize=4.0in \epsfbox{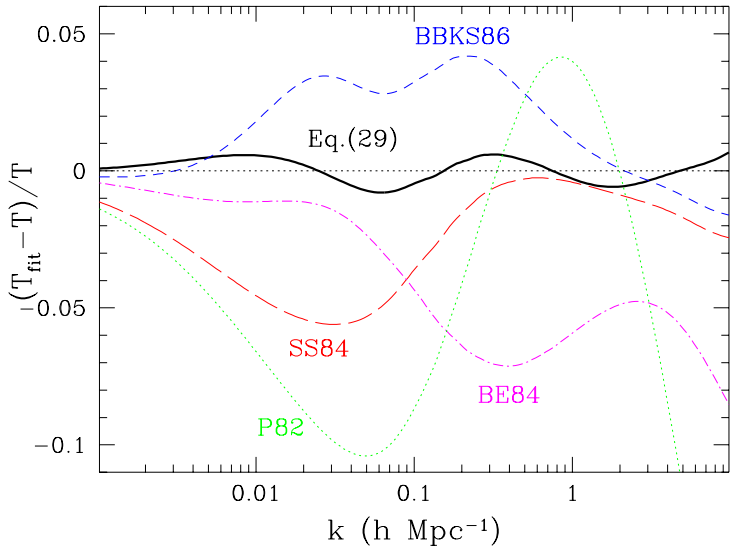}
\end{center}
\caption{Comparison of various zero baryon transfer functions to the
numerical calculation of CMBfast (\protect\cite{Sel96}\ 1996).  The
fitting form of equation (\protect\ref{eq:T0}) agrees with CMBfast to
$1\%$, whereas previous approximations \protect\cite{Pee82}\ (1982) [P82],
\protect\cite{Sta84}\ (1984) [SS84], and \protect\cite{Bar86}\ (1986) [BBKS86].
Also shown is the form of \protect\cite{Bon84}\ (1984) [BE84], a 3\% baryon 
fraction model that has often been used as a zero-baryon proxy.}
\label{fig:T0}
\end{figure}

The following functional form satisfies 
these criteria and is a {\it better} fit to the zero baryon case
extrapolated from trace-baryon models calculated by CMBfast
\begin{eqnarray}
T_0(q) &=& {L_0 \over {L_0 + C_0 q^2}}, \nonumber\\
\label{eq:T0}
L_0(q) &=& \ln(2e + 1.8q), \\
C_0(q) &=& 14.2 +{731 \over 1+62.5 q}. \nonumber 
\end{eqnarray} 
Note that this form is not only more accurate than the \cite{Bar86}\ 
(1986) one but is also simpler: there are fewer parameters and the
coefficients $1.8$ and $14.2$ are derived theoretically.  The parameter
$731$, the small-$q$ quadratic deviation, has been fit rather than
derived to account for the small correction due to anisotropic stress
gradients. In Fig.~\ref{fig:T0} we show a comparison of this form to
numerical calculations and various fitting formulae in the literature.
Our formula agrees with numerical calculations at the same level as
different numerical calculations agree with each other (\cite{Hu95}
1995), i.e.~to 1\% through the CMB and large-scale structure regimes.

The presence of baryons has commonly been included by fitting 
a constant shape parameter $\Gamma$ (\cite{Pea94}\ 1994;
\cite{Sug95}\ 1995).  That such an approach can work on small scales
can be seen as follows.  On small scales, the effect of the baryons
is a constant suppression by the factor $\alpha_c \Omega_c/\Omega_0$.
Since the transfer function there is proportional to $(k/\Gamma_0)^{-2}$, 
a rescaling of $\Gamma_0$ approximates this effect.  However, this
simple rescaling of $\Gamma_0$ does not properly treat the region 
observable through large-scale structure.  Above the sound horizon,
the baryons and CDM are indistinguishable, and so the transfer
function is close to that parameterized by 
$\Gamma_0 \equiv \Omega_0 h$.  Below the horizon,
if one neglects the oscillations caused by the baryons, the transfer
function is suppressed and roughly follows that of a rescaled $\Gamma$.
Hence the transition around the sound horizon cannot be fit by a 
single $\Gamma$.

\begin{figure}[bt]
\begin{center} \leavevmode \epsfxsize=4.0in \epsfbox{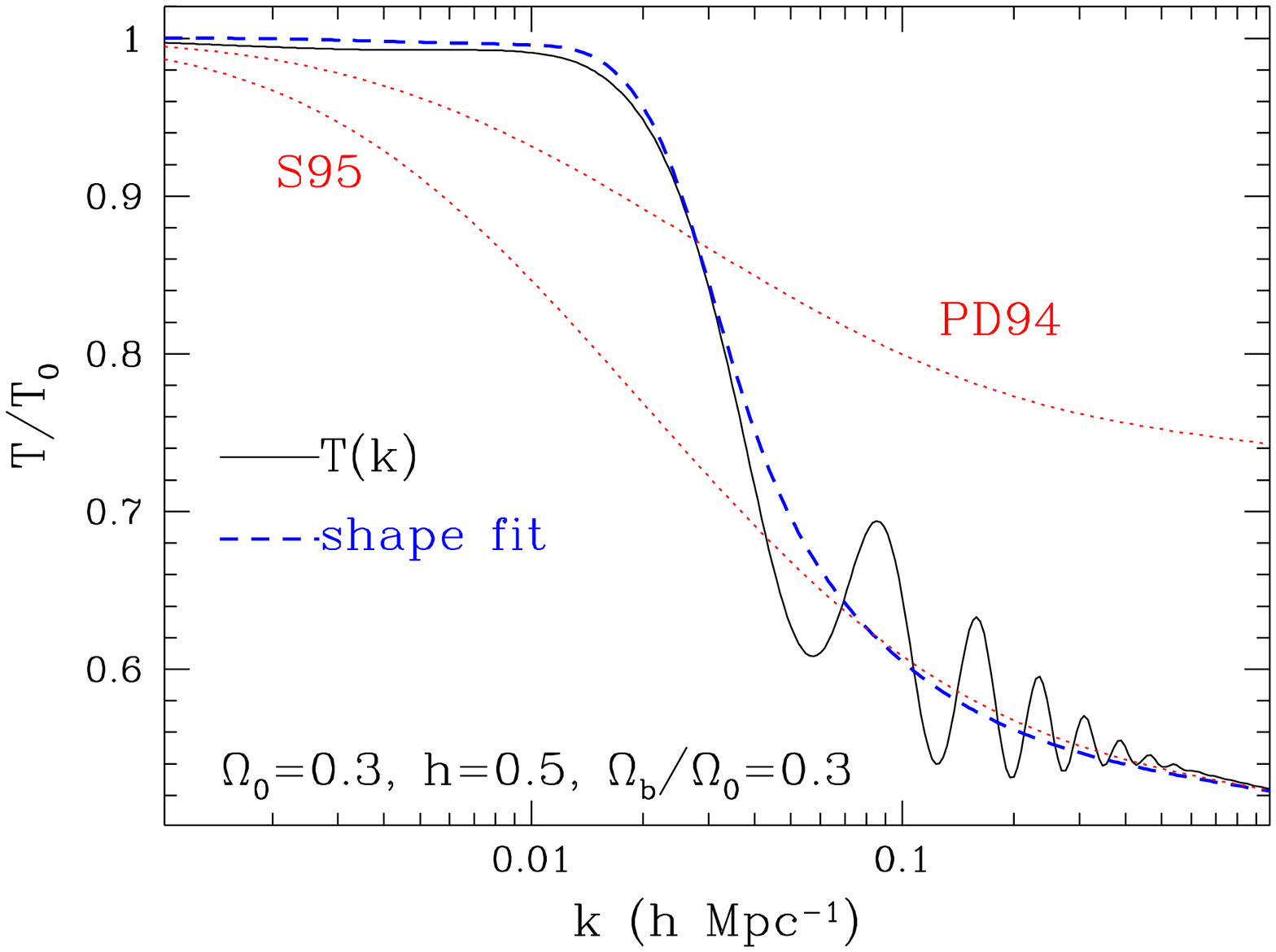}
\end{center}
\caption{Distortions to the shape of the transfer function as measured
by the true transfer function over the zero baryon form. The shape
function of equation (\protect\ref{eq:gammaeff}) adequately describes the 
true curve apart from the oscillations, whereas the constant shift in $\Gamma$
described by \protect\cite{Pea94}\ (1994) [PD94] and 
\protect\cite{Sug95}\ (1995) [S95] do not.}
\label{fig:gamma}
\end{figure}

A reasonable fit to the non-oscillatory part of the transfer function
can be written by rescaling $\Gamma_{\rm eff}(k)$ as one moves through
the sound horizon
\begin{eqnarray}
\Gamma_{\rm eff}(k) &=& \Omega_0 h \left(\alpha_\Gamma + 
{1-\alpha_\Gamma\over 1+(0.43ks)^4}\right), \label{eq:gammaeff}\\
\alpha_\Gamma &=& 1 - 0.328\ln(431\Omega_0h^2)\, {\Omega_b \over \Omega_0} +
0.38\ln(22.3 \Omega_0h^2)\, \left( {\Omega_b\over \Omega_0} \right)^2. 
\label{eq:alphagamma}
\end{eqnarray}
Defining $q_{\rm eff}$ as in equation (\ref{eq:Gamma}), we find that $T(k)
\approx T_0(q_{\rm eff})$.  Figure \ref{fig:gamma} shows an example.
Note the simpler form of $s$ in equation (\ref{eq:sfit}) may be used here.
$\alpha_\Gamma$ is nearly $\sqrt{\alpha_c\Omega_c/\Omega_0}$, the
radicand of which is plotted in Fig.\ \ref{fig:suppression}; we
provide the above form for simplicity and to account for small
deviations at the higher $\Omega_0 h^2$ values.  The latter arise
because the fit has been optimized in the observable region
$k\lesssim0.1 {\rm\,Mpc}^{-1}$, where the CDM transfer function is not
quite in its $k^{-2}$ small-scale limit.  Of course, neglecting the
oscillatory contributions is not a good approximation for
$\Omega_b/\Omega_0 \simgt 0.5$.  In addition to the obvious omission of
the wiggles, we have fit $\alpha_\Gamma$ to the transfer function
$T(k)$ and not the observable $T(k)^2$.  This neglects the power
introduced by the square of the oscillatory term.

We caution the reader that the small-scale asymptote $\Gamma_{\rm eff}
= \alpha_\Gamma \Omega_0 h$ arises from the normalization rather than
shape and therefore should not be conflated with $\Gamma$ derived from
redshift surveys (\cite{Pea94}\ 1994).  Pure normalization distinctions are
not observable with current redshift surveys; instead, one estimates
$\Gamma$ by fitting power spectra of arbitrary normalization and
relying on the differences in shape and power-law slope introduced by
the variation in $\Gamma$.  However, when combined with the COBE
normalization (e.g.\ \cite{Bun97}\ 1997), the normalized predictions 
from equations (\ref{eq:gammaeff}) and (\ref{eq:alphagamma}) are appropriate.  

However, equation (\ref{eq:gammaeff}) and Fig.\ \ref{fig:gamma} do make clear
that a simple rescaling of $\Gamma$ is not accurate near the sound
horizon.  Future measurements of large-scale structure expected from
the Sloan Digital Sky Survey (\cite{Gun95}\ 1995) and 2-degree Field 
(\cite{Tay95}\ 1995) surveys will be 
sufficiently precise as to detect the deviations from a single $\Gamma$
model (\cite{Teg97}\ 1997; \cite{Gol97}\ 1997).  A key signature is that
the bend from $T\approx1$ to $T\propto k^{-2}$ becomes sharper as
baryons are introduced.  Detecting this change will require that scales
around $s/\pi$, c.f.\ equation (\ref{eq:sfit}), are well-observed.

\section{Discussion}
We have presented an accurate, well-motivated fitting form for the
transfer function of a general CDM--baryon universe.  The formula is
generally accurate to better than 10\% in fully baryonic universes and
better than 5\% in cosmologies with $\Omega_b/\Omega_0\simlt0.5$.
While the available numerical codes will provide yet more accurate
transfer functions, the fitting form here should be useful for
characterizing trends in cosmological parameter space.  Moreover, by
separating the various physical aspects in the analytic form, we hope
to provide physical intuition for the effects, their 
interrelationships, and their correspondence with analogous effects in
the CMB. 

As applications of the form, we gave quantitative assessments of the
location and amplitude of the baryonic features in the linear 
regime.  We quantified the
suppression of power on scales below the sound horizon due to the
admixture of baryons and showed that this suppression is not well-fit
by a rescaling of $\Gamma$ if one is probing scales near the sound
horizon.  An alternative model, based on an interpolation in $\Gamma$,
gives a reasonable fit to the non-oscillatory portion of the transfer
function provided that $\Omega_b/\Omega_0\simlt \Omega_0 h^2 + 0.2$.
Finally, we gave a new fitting form for the zero-baryon limit that is
more accurate on small scales than those used previously.
A summary of how to use these various formulae is given in Appendix A.

Baryonic features, like the acoustic peaks in the CMB
(\cite{Hu96b}\ 1996),
transcend the adiabatic CDM paradigm discussed in
this paper.  In fact any gravitational instability model where
reionization took place no earlier than $z_d \sim 270 (\Omega_0 h^2)^{1/5}$,
the Compton drag epoch for a fully ionized universe, must possess acoustic
effects at some level.  Evidence of these effects is strong indication that
fluctuations were generated in the early universe.  Moreover, when combined with
CMB observations, baryonic features allow strong consistency tests 
for the predicted growth of fluctuations as both 
sets of features reflect the underlying acoustic oscillations
before recombination.  A measurement of the sound horizon from the matter
power spectrum combined with its angular extent from the CMB would allow
an angular diameter distance test for curvature in the
universe that is largely free of cosmological assumptions.  
Baryonic features in the matter power spectrum thus represent a valuable
resource for cosmological information, but one that may be difficult to 
mine observationally.

\noindent{\it Acknowledgments:} 
The CMBfast package 
(http://arcturus.mit.edu:80/$\sim$matiasz/CMBFAST/cmbfast.html)
was used to generate numerical transfer functions.  
We thank S.\ Boughn, U.\ Seljak, J.\ Silk, D.\ Spergel, 
A.\ Szalay, M.\ Tegmark, and M.\ White for useful discussions. 
W.H.\ acknowledges support from the W.M. Keck foundation and the
hospitality of the Aspen Center for Physics.
D.J.E.\ acknowledges support from NSF PHY-9513835.

\appendix
\section{A User's Guide}
In this paper, we have presented a fitting form for the transfer function
in a CDM and baryon cosmology with adiabatic perturbations.  The formula
is given as equations (\ref{eq:zeq})--(\ref{eq:ksilk}), 
(\ref{eq:q})--(\ref{eq:betac}), and (\ref{eq:alphab})--(\ref{eq:betab}).
Alternatively, if one prefers a simpler form that accurately 
represents the baryon-induced suppression on intermediate scales but
that ignores the acoustic oscillations in the transfer function,
one may use the form given in equations (\ref{eq:sfit}) and
(\ref{eq:Gamma})--(\ref{eq:alphagamma}). 
The oscillations are fairly small ($\lesssim\!20\%$ modulation in power)
for $\Omega_b/\Omega_0 \lesssim \Omega_0 h^2 +0.2$, so in these cases 
the simpler form will be appropriate for many applications.  We also provide
in equation (\ref{eq:T0}) a more accurate form for the zero-baryon 
transfer function.  Electronic versions of the formulae in this paper
may be found at
\begin{center}
http://www.sns.ias.edu/$\sim$whu/transfer/transfer.html.
\end{center}

The power spectrum of the density fluctuations is then proportional
to the initial power spectrum times the square of the transfer function.
In the most usual case, the initial power spectrum is taken to be
a power-law, so that $P(k)\propto k^n T^2(k)$, where $n=1$ is the familiar
Harrison-Zel'dovich-Peebles scale-invariant case.  

While the transfer function is independent of late-time effects such
as the presence of a cosmological constant or curvature, the 
magnitude and time-dependence of the normalization of the power
spectrum does depend on these effects.  \cite{Bun97} (1997) calculate
the present-day normalization of the power spectrum implied
by the 4-year COBE anisotropy measurement and provide the following
fitting forms for flat and open cosmologies:
\begin{eqnarray}
\left.\Delta^2(k)\right|_{z=0} &\equiv& {k^3\over 2\pi^2} P(k) = 
\delta_H^2\left(ck\over H_0\right)^{3+n}T^2(k),\\
\delta_H &=& 1.95\times10^{-5} \Omega_0^{-0.35-0.19\ln\Omega_0-0.17\tilde{n}}
e^{-\tilde{n}-0.14\tilde{n}^2}\hspace{24pt}(\Lambda=0),\\
\delta_H &=& 1.94\times10^{-5} \Omega_0^{-0.785-0.05\ln\Omega_0} 
e^{-0.95\tilde{n}-0.169\tilde{n}^2}\hspace{24pt}(\Lambda=1-\Omega_0),
\end{eqnarray}
where $\tilde{n}=n-1$ and the contribution of tensor perturbations to
the observed anisotropies has been assumed to be zero (see \cite{Bun97}
1997 for more details).  The $1\sigma$ statistical uncertainty is 7\%,
and the error in the above fits are much smaller than this for
$0.2\le\Omega_0\le1$ and $0.7\le n\le1.2$.

To extend this normalization to earlier times, one needs to scale the
power spectrum by the square of the growth function $D_1(z)$ (Peebles 1980).  
As is well-known, for $\Omega_0=1$, $D_1 = (1+z)^{-1}$ at redshift $z$.
For other cosmologies, one can use the approximation 
(\cite{Lah91}\ 1991; \cite{Car92} 1992)
\begin{equation}
D_1(z) = (1+z)^{-1}{5\Omega(z)\over2}\left\{\Omega(z)^{4/7}
-\Omega_\Lambda(z)+[1+\Omega(z)/2][1+\Omega_\Lambda(z)/70]\right\}^{-1}.
\end{equation}
Here $\Omega(z)$ and $\Omega_\Lambda(z)$ are the density parameters
as seen by an observer at redshift $z$; hence
\begin{eqnarray}
\Omega(z) &=& {\Omega_0 (1+z)^3\over 
\Omega_\Lambda+\Omega_R (1+z)^2+\Omega_0(1+z)^3},\\
\Omega_\Lambda(z) &=& {\Omega_\Lambda \over 
\Omega_\Lambda+\Omega_R (1+z)^2+\Omega_0(1+z)^3},
\end{eqnarray}
where $\Omega_0$ is the matter density in units of critical,
$\Omega_\Lambda=\Lambda/3H_0^2$ represents the cosmological constant
$\Lambda$, and $\Omega_R = 1-\Omega_0-\Omega_\Lambda$ represents the
effects of curvature.  

The normalization of the power spectrum is such that the variance
of mass fluctuations inside a sphere of radius $R$ is 
\begin{equation}
\sigma^2(R) = \int_0^\infty {dk\over k}\Delta^2(k)
\left(3j_1(kR)\over kR\right)^2,
\end{equation}
where $j_1(x)=(x\,\cos x-\sin x)/x^2$.


\end{document}